\pdfoutput=1
\RequirePackage{ifpdf}
\ifpdf 
\documentclass[pdftex]{sigma}
\else
\documentclass{sigma}
\fi

\numberwithin{equation}{section}
\newtheorem{Theorem}{Theorem}[section]
\newtheorem{Corollary}[Theorem]{Corollary}
\newtheorem{Proposition}[Theorem]{Proposition}
 { \theoremstyle{definition}
\newtheorem{Remark}[Theorem]{Remark} }

\usepackage{srcltx}

\DeclareMathOperator{\Ann}{Ann}
\DeclareMathOperator{\Spin}{Spin}
\DeclareMathOperator{\Res}{Res}

\begin{document}

\allowdisplaybreaks

\renewcommand{\thefootnote}{$\star$}

\renewcommand{\PaperNumber}{007}

\FirstPageHeading

\ShortArticleName{The $(n,1)$-Reduced DKP Hierarchy}

\ArticleName{The $\boldsymbol{(n,1)}$-Reduced DKP Hierarchy,\\
the String Equation and $\boldsymbol{W}$ Constraints\footnote{This paper is a~contribution to the Special
Issue in honor of Anatol Kirillov and Tetsuji Miwa.
The full collection is available at
\href{http://www.emis.de/journals/SIGMA/InfiniteAnalysis2013.html}
{http://www.emis.de/journals/SIGMA/InfiniteAnalysis2013.html}}}

\Author{Johan VAN DE LEUR}

\AuthorNameForHeading{J.~van de Leur}

\Address{Mathematical Institute, University of Utrecht,\\
P.O.~Box 80010, 3508 TA Utrecht, The Netherlands}

\Email{\href{mailto:J.W.vandeLeur@uu.nl}{J.W.vandeLeur@uu.nl}}
\URLaddress{\url{http://www.staff.science.uu.nl/~leur0102/}}

\ArticleDates{Received September 23, 2013, in f\/inal form January 09, 2014; Published online January 15, 2014}

\Abstract{The total descendent potential of a~simple singularity satisf\/ies the Kac--Waki\-moto principal
hierarchy. Bakalov and Milanov showed recently that it is also a~highest weight vector for the corresponding
$W$-algebra. This was used by Liu, Yang and Zhang to prove its uniqueness.
We construct this principal hierarchy of type~$D$ in a~dif\/ferent way, viz.\
as a~reduction of some DKP hierarchy.
This gives a~Lax type and a~Grassmannian formulation of this hierarchy.
We show in particular that the string equation induces a~large part of the $W$ constraints of Bakalov and Milanov.
These constraints are not only given on the tau function, but also in terms of the Lax and Orlov--Schulman operators.}


\Classification{17B69; 17B80; 53D45; 81R10}

\renewcommand{\thefootnote}{\arabic{footnote}}
\setcounter{footnote}{0}

\section{Introduction}

Givental, Milanov, Frenkel, and Wu, showed a~in a~series of publications~\cite{FGM,G, GM, Wu1} that the
total descendant potential of an~$A$, $D$ or~$E$ type singularity satisf\/ies the Kac--Wakimoto
hierarchy~\cite{KW}.
Recently Bakalov and Milanov showed in~\cite{BM} that this potential is also a~highest weight vector for
the corresponding $W$-algebra.
For type $A$ Fukuma, Kawai and Nakayama~\cite{FKN} showed that these $W$ constraints can be obtained
completely from the string equation.
This was used by Kac and Schwarz~\cite {KS} to show that this $A_n$ potential is a~unique $(n+1)$-reduced KP
tau function, if one assumes that it corresponds to a~point in the big cell of the Sato Grassmannian.

Uniqueness for type $D$ and $E$ singularities, together with the $A$ case as well, was recently shown by Liu,
Yang and Zhang in~\cite{LYZ}.
They use the results of~\cite{BM} and the twisted vertex algebra construction to obtain this result.
Both constructions use the Kac--Wakimoto principal hierarchy construction of~\cite{KW}.

In this paper we obtain the principal realization of the basic module of type $D_{n}^{(1)}$ as a~certain
reduction of a~representation of $D_\infty$.
The reduction of the corresponding DKP-type (or sometimes also called 2-component BKP) hierarchy gives
Hirota bilinear equations for the corresponding tau functions.
This gives an equivalent but slightly dif\/ferent formulation of Kac--Wakimoto $D_n$ principal
hierarchy~\cite{KW}.
The total descendent potential of a~$D_n$ type singularity satisf\/ies these equations.
This approach has 3 advantages: (1)~there is a~Lax type formulation for this hierarchy; (2)~there is
a~Grassmannian formulation for this reduced hierarchy; (3)~one can show that the string equation generates
part of the $W$-algebra constraints.
This makes it possible to describe~-- at least part of~-- the
$W$-algebra constraints in terms of
pseudo-dif\/ferential operators and in terms of the corresponding Grassmannian.

This approach, viz.\
obtaining the principal hierarchy of type $D$ as a~reduction of the 2-com\-ponent BKP hierarchy,
which describes the $D_\infty$-group orbit of the highest weight vector,
was also considered by Liu, Wu and Zhang in~\cite{LWZ}.
They even obtain Lax equations.
However, their Lax equations are formulated dif\/ferently than the ones in this paper.
They use certain (scalar) pseudo-dif\/ferential operators of the second type, where we need not only the
basic representation of type $D$, but also the other level one module.
As such we obtain a~pair of tau functions $\tau_0$ and $\tau_1$, which are related.
The equations on both tau functions provide $(2\times 2)$-matrix pseudo-dif\/ferential operators, with which
we can formulate a~slightly dif\/ferent, but probably equivalent, Lax equation.
However, in both approaches the equations on the tau-function $\tau_0$ is the same.
Wu~\cite{Wu2} used the approach of~\cite{LWZ} to study the Virasoro-constraints, he showed that they can be
obtained from the string equation.
Using the $(2\times 2)$-matrix pseudo-dif\/ferential approach of this paper, we recover Wu's result and even
more, the string equation not only produces the Virasoro constraints but even produces a~large part of the
Bakalov--Milanov~\cite{BM} $W$ constraints, but not all.

\section[The $(n,1)$-reduced DKP hierarchy]{The $\boldsymbol{(n,1)}$-reduced DKP hierarchy}

\subsection[The principal hierarchy for $D_{n+1}^{(1)}$]{The principal hierarchy for $\boldsymbol{D_{n+1}^{(1)}}$}

The principal hierarchy of the af\/f\/ine Lie algebra $D_{n+1}^{(1)}$ can be described in many dif\/ferent
ways~\cite{K,KW}.
Here we take the approach of {ten Kroode} and the author~\cite{tKvdL}
and describe this hierarchy as
a~reduction of the 2-component BKP hierarchy, i.e., we introduce two neutral or twisted fermionic f\/ields
and obtain a~representation of the Lie algebra of $d_\infty$.
We def\/ine an equation which describes the corresponding $D_\infty$ group orbit of the highest weight
vector.
Following Jimbo and Miwa~\cite{JM} we use a~certain reduction procedure, which reduces the group to
a~smaller group, viz., to the group corresponding to $D_{n+1}^{(1)}$ in its principal realization and thus
obtain a~larger set of equations for elements in the group orbit.
\begin{Remark}
\label{r1}
It is important to note the following.
The Kac--Wakimoto principal hierarchy of type $D_{n+1}^{(1)}$ characterizes the group orbit of the highest
weight vector of type $D_{n+1}^{(1)}$ in the principal realization (see~\cite[Theorem 0.1]{KW} or~\cite{K}).
Jimbo and Miwa show in~\cite{JM} that elements of this group orbit satisfy this $D_{n+1}^{(1)}$ reduction
of this DKP or 2 component BKP hierarchy.
Since the total descendent potential of a~$D_{n+1}$ singularity satisf\/ies the Kac--Wakimoto hierarchy it
is an element in this $D_{n+1}^{(1)}$ group orbit and hence also satisf\/ies this Jimbo--Miwa
$D_{n+1}^{(1)}$ principal reduction or $(n,1)$-reduced DKP hierarchy.
\end{Remark}
Let $n$ be a~positive integer, consider the following Clif\/ford algebra ${\rm Cl}(\mathbb{C}^\infty) $ on the
vector space~$\mathbb{C}^\infty$ with basis $\phi_{\frac{i}{2n}}^1$, $ \phi_{\frac{i}{2}}^2$, with
$i\in\mathbb{Z}$ and symmetric bilinear form
\begin{gather*}
\left(\phi_{\frac{i}{2n}}^1,\phi_{\frac{j}{2n}}^1\right)
=\left(\phi_{\frac{i}{2}}^2,\phi_{\frac{j}{2}}^2\right)=(-)^i\delta_{i,-j},
\qquad
\left(\phi_{\frac{i}{2n}}^1,\phi_{\frac{j}{2}}^2\right)=0.
\end{gather*}
The Clif\/ford algebra has the usual commutation relations:
\begin{gather*}
\phi_{\frac{i}{2n}}^1\phi_{\frac{j}{2n}}^1+\phi_{\frac{j}{2n}}^1\phi_{\frac{i}{2n}}^1=(-)^i\delta_{i,-j}
=\phi_{\frac{i}{2}}^2\phi_{\frac{j}{2}}^2+\phi_{\frac{j}{2}}^2\phi_{\frac{i}{2}}^2,
\qquad
\phi_{\frac{i}{2n}}^1\phi_{\frac{j}{2}}^2+\phi_{\frac{j}{2}}^2\phi_{\frac{i}{2n}}^1=0.
\end{gather*}
We def\/ine its corresponding Spin module $V$ with vacuum vector $|0\rangle$ as follows (cf.\
\cite{tKvdL}):
\begin{gather*}
\phi_{\frac{i}{2n}}^1|0\rangle=\phi_{\frac{i}{2}}^2|0\rangle=0,
\quad
i>0,
\qquad
\left(\phi_{0}^1+i\phi_{0}^2\right)|0\rangle=0.
\end{gather*}
The normal ordered elements $:\phi_{i}^a\phi_{j}^b:$ form the inf\/inite Lie algebra of type $d_\infty$,
where the central elements acts as $1$, see~\cite{tKvdL} for more details.
The best way to describe the af\/f\/ine Lie algebra~$D_{n+1}^{(1)}$ is to introduce, following~\cite{BK},
$\omega=e^{\frac{\pi i}n}$ the $2n$-th root of~1, and write
\begin{gather*}
\varphi(z)=\sum_{m\in\frac{1}{2n}\mathbb{Z}}\varphi_{(m)}z^{-m-1},
\qquad
\text{then}
\qquad
\varphi(e^{2\pi i k}z)=\sum_{m\in\frac{1}{2n}\mathbb{Z}}\omega^{-2kmn}\varphi_{(m)}z^{-m-1}.
\end{gather*}
The f\/ields corresponding to the elements in the Clif\/ford algebra are
\begin{gather*}
\phi^{1}(z)=\sum_{i\in\mathbb{Z}}\phi_{\frac{i}{2n}}^{1}z^{\frac{-n-i}{2n}},
\qquad
\phi^{2}(z)=\sum_{i\in\mathbb{Z}}\phi_{\frac{i}{2}}^{2}z^{\frac{-1-i}{2}}.
\end{gather*}
Then the commutation relations can be described as follows in term of the anti-commutator $\{\, ,\}$
\begin{gather*}
\left\{\phi^{1}(z),\phi^{1}\big(e^{2\pi i n}w\big)\right\}=(-)^n\sum_{j=0}^{2n-1}\delta_j(z-w),
\qquad
\left\{\phi^{2}(z),\phi^{2}\big(e^{2\pi i}w\big)\right\}=-\sum_{j=0}^{1}\delta_{jn}(z-w),
\\
\left\{\phi^{1}(z),\phi^{2}(w)\right\}=0,
\end{gather*}
where $\delta_j(z-w)$ is the $2n$-twisted delta function, e.g.~\cite{BK}:
\begin{gather*}
\delta_j(z-w)=z^{\frac{j}{2n}}w^{\frac{-j}{2n}}\delta(z-w)=\sum_{k\in\frac{j}{2n}+\mathbb{Z}}z^kw^{-k-1}.
\end{gather*}

Then, see~\cite{tKvdL}, the modes of the f\/ields
\begin{gather*}
:\phi^{a}(e^{2\pi i k}z)\phi^{b}\big(e^{2\pi i\ell}z\big):,
\qquad
1\le a,b\le2, \qquad  0\le k,\ell\le2n-1,
\end{gather*}
together with 1 span the af\/f\/ine Lie algebra of type $D_{n+1}^{(1)}$ in its principal realization.
The spin module $V$ splits in the direct sum of two irreducible components when restricted to
$D_{n+1}^{(1)}$.
The irreducible components $V=V_0$ and $V_1$ correspond to the $\mathbb{Z}_2$ gradation given by
\begin{gather*}
\deg|0\rangle=0,
\qquad
\deg\phi^{\pm a}_k=1.
\end{gather*}
The highest weight vector of $V_0$ is $|0\rangle$, the highest weight vector of $V_1$ is
\begin{gather*}
|1\rangle=\frac{1}{\sqrt2}\big(\phi_{0}^1-i\phi_{0}^2\big)|0\rangle.
\end{gather*}
Here $V_0$ is the basic representation, $V_1$ is an other level 1 module.
Both modules are isomorphic.

\subsection{The DKP hierarchy and its principal reduction}

The DKP hierarchy is the following equation on ${\mathfrak T}\in V_0$:
\begin{gather*}
\Res_z\left((-)^n\phi^{1}(z){\mathfrak T}\otimes\phi^{1}\big(e^{2\pi i n}z\big){\mathfrak T}-\phi^{2}
(z){\mathfrak T}\otimes\phi^{2}\big(e^{2\pi i}z\big){\mathfrak T}\right)=0.
\end{gather*}
This equation describes an element in the $D_\infty$-group orbit of $|0\rangle$.

If one restricts the action on $|0\rangle$ to the loop group of type $D_{n+1}^{(1)}$, the orbit is smaller
and is given by more equations.
The principal reduction, of~\cite{DJKMEuclidian, JM} induces the following.
If ${\mathfrak T}\in V_0$ is in this loop group orbit of $|0\rangle$, it satisf\/ies the $(n,1)$-reduced
DKP hierarchy for all integers $p\ge0$
\begin{gather}
\Res_zz^p\left((-)^n\phi^{1}(z){\mathfrak T}\otimes\phi^{1}\big(e^{2\pi i n}z\big){\mathfrak T}-\phi^{2}
(z){\mathfrak T}\otimes\phi^{2}\big(e^{2\pi i}z\big){\mathfrak T}\right)=0.
\label{redDKP}
\end{gather}

However, for us it will be more convenient not only to use the action on $|0\rangle$ but also on
$|1\rangle$ and write ${\mathfrak T}_a$ for the action of the loop group on $|a\rangle$, where $a=0,1$.
One thus obtains
\begin{gather}
\label{redDKP2}
\Res_zz^p\left((-)^n\phi^{1}(z){\mathfrak T}_a\otimes\phi^{1}\big(e^{2\pi i n}z\big){\mathfrak T}_b-\phi^{2}
(z){\mathfrak T}_a\otimes\phi^{2}\big(e^{2\pi i}z\big){\mathfrak T}_b\right)=\delta_{a+b,1}\delta_{p0}{\mathfrak T}
_b\otimes{\mathfrak T}_a
\end{gather}
for all integers $p\ge 0$, here $a,b=0,1$.

\subsection{A Grassmannian description}

We follow the description of~\cite{KL2}.
The Clif\/ford algebra ${\rm Cl}(\mathbb{C}^\infty) $ has a~natural $\mathbb{Z}_2$-gradation
${\rm Cl}(\mathbb{C}^\infty)={\rm Cl}_0(\mathbb{C}^\infty)\oplus {\rm Cl}_1(\mathbb{C}^\infty)$, where
${\rm Cl}_0(\mathbb{C}^\infty)_0$ consists of products of an even number of elements from $\mathbb{C}^\infty$.
Let $\Spin(\mathbb{C}^\infty)$ denote the multipicative group of invertible elements in $a\in
{\rm Cl}_0(\mathbb{C}^\infty)$ such that $a\mathbb{C}^\infty a^{-1}=\mathbb{C}^\infty$.
There exists a~homomorphism $T:\Spin(\mathbb{C}^\infty)\to D_\infty$ such that $T(g)(v)=gvg^{-1}$.
Thus $T(g)$ is orthogonal, i.e., $(T(g)(v),T(g)(w))=(v,w)$, in fact it is an element in
${\rm SO}(\mathbb{C}^\infty)$.
Let $a=0,1$, then
\begin{gather*}
\Ann(g|a\rangle)=\{v\in\mathbb{C}^\infty|vg|a\rangle=0\}
=\big\{gvg^{-1}\in\mathbb{C}^\infty|v|a\rangle=0\big\}=T(g)\big(\Ann(|a\rangle)\big).
\end{gather*}
Since
\begin{gather}
\label{vacuum}
\Ann(|a\rangle)=\mathbb{C}\frac{\phi_{0}^1+(-)^a i\phi_{0}^2}{\sqrt2}\oplus\bigoplus_{i>0}\mathbb{C}
\phi_{\frac{i}{2n}}^1\oplus\mathbb{C}\phi_{\frac{i}{2}}^2,
\end{gather}
it is easy to verify that $\Ann(|a\rangle)$ for $a=0,1$ is a~maximal isotropic subspace of
$\mathbb{C}^\infty$ and hence $\Ann(g|a\rangle)$ for $a=0,1$ and $g\in \Spin(\mathbb{C}^\infty)$ is
also maximal isotropic.
Hence an element in the $D_\infty$ group orbit of the vacuum vector produces two unique maximal isotropic
subspaces.
We can say even more, the modif\/ied DKP hierarchy, i.e.
equation~\eqref{redDKP2} with $p=0$ and $\{ a,b\}=\{ 0,1\}$, has the following geometric interpretation,
see also~\cite{KL2} for more information,
\begin{gather*}
\dim\left(\Ann(g|a\rangle-\Ann(g|b\rangle)\right)=1,
\qquad
0\le a\ne b\le1.
\end{gather*}
Note that this follows immediately from~\eqref{vacuum}.
Let $e_1$ and $e_2$ be the orthonormal basis of $\mathbb{C}^2$ we identify
\begin{gather}
\label{phit}
\phi_{\frac{i}{2n}}^1=t^{\frac{i}{2n}}e_1,
\qquad
\phi_{\frac{i}{2}}^2=t^{\frac{i}{2}}e_2,
\end{gather}
where we assume that the bilinear form does not change, i.e.,
\begin{gather*}
\big(t^{\frac{i}{2n}}e_1,t^{\frac{j}{2n}}e_1\big)=(-)^i\delta_{i,-j},
\qquad
\big(t^{\frac{i}{2}}e_2,t^{\frac{j}{2}}e_2\big)=(-)^i\delta_{i,-j},
\qquad
\big(t^{\frac{i}{2n}}e_1,t^{\frac{j}{2}}e_2\big)=0.
\end{gather*}
We think of $t=e^{i\theta}$ as the loop parameter.
Now if $g$ corresponds to an element in $D_{n+1}^{(1)}$, then $\Ann(g|a\rangle)$ satisf\/ies
\begin{gather*}
t\Ann(g|a\rangle)\subset\Ann(g|a\rangle),
\qquad
a=0,1.
\end{gather*}

\subsection{A bosonization procedure}

In general there are many dif\/ferent bosonizations for the same level one $D_{n+1}^{(1)}$ module
(see~\cite{KP} and~\cite{tKvdL}).
Kac and Peterson~\cite{KP} showed
that for every conjugacy class of the Weyl group of type $D_{n+1}$ there
is a~dif\/ferent realization.
The principal realization f\/irst obtained in~\cite {KKLW} is the realization which is connected to
a~Coxeter element in the Weyl group (all Coxeter elements form one conjugacy class).
As such the bosonization procedure for this principal realization is unique and well known, see, e.g.,~\cite{tKvdL}.
Here we do not take the usual one, but the one which is related to the $D_{n+1}$ singularities as in the
paper of Bakalov and Milanov~\cite{BM}.
This means that we introduce a~parameter $\sqrt \hbar$ and that we choose the realization of the Heisenberg
algebra slightly dif\/ferent from the usual one.

The bosonization of this principal hierarchy consists of identifying $V$ with the space
$F=\mathbb{C}[\theta, q_k^a;a=1,2,\dots, n+1, k=0,1,\ldots]$.
Here $\theta$ is a~Grassmann variable satisfying $\theta^2=0$.
Let~$\sigma$ be the isomorphism that maps $V$ into $F$,
we take $\sigma(V_0)=F_0=\mathbb{C}[q_k^a;a=1,2,\dots, n+1, k=0,1,\ldots]$
and $\sigma(V_1)=F_1=\theta\mathbb{C}[ q_k^a;a=1,2\dots n+1,k=0,1,\ldots]$.
The Heisenberg algebra, $\alpha_k^a$ is def\/ined by
\begin{gather*}
\alpha^1(z)=\sum_{i\in\frac{1}{2n}+\frac{1}{n}\mathbb{Z}}\alpha_i^1z^{-i-1}:=\frac{(-1)^n}{2\sqrt n}
:\phi^{1}(z)\phi^{1}\big(e^{2\pi i n}z\big):,
\\
\alpha^2(z)=\sum_{i\in\frac{1}{2}+\mathbb{Z}}\alpha_i^2z^{-i-1}:=\frac{-1}{2}:\phi^{2}(z)\phi^{2}
\big(e^{2\pi i}z\big):.
\end{gather*}
Then
\begin{gather*}
\big[\alpha^a_k,\alpha^b_\ell\big]=k\delta_{ab}\delta_{k,-\ell}
\qquad
\text{and}
\qquad
\big[\alpha^1_k,\phi^1(z)\big]=\frac{z^k}{\sqrt n}\phi^1(z),
\qquad
\big[\alpha^2_k,\phi^2(z)\big]={z^k}\phi^1(z).
\end{gather*}
\begin{Remark}
\label{relBM}
Note that in the notation of~\cite{BM}, $n=N-1$,
\begin{gather*}
\alpha^1(z)=Y(\sqrt n v_1,z)=\sqrt n Y(v_1,z),
\qquad
\alpha^2(z)=Y(v_{n+1},z)
\end{gather*}
and
\begin{gather}
\label{factors}
\phi^1(z)=\frac{1}{\sqrt{2n}}Y\big(e^{v_1},z\big),
\qquad
\phi^2(z)=\frac{1}{\sqrt{2}}Y\big(e^{v_{n+1}},z\big).
\end{gather}
Here the $v_i$ form an orthonormal basis of the Cartan subalgebra of the Lie algebra of type $D_{n+1}$.
Elements $e^{v_i}$ are elements in the group algebra of the root lattice of type $B_{n+1}$, which has as
basis the elements~$v_i$.
This construction is related to an automorphism $\rho$, which is a~lift of a~Coxeter element in the Weyl
group and which gives the Kac--Peterson twisted realization~\cite{KP}, see also~\cite{tKvdL} for more details.
$\rho$ acts on the $v_1,v_2,\ldots,v_n,v_{n+1}$ as follows
\begin{gather*}
v_1\mapsto v_2\mapsto\cdots\mapsto v_n\mapsto-v_1,
\qquad
v_{n+1}\mapsto-v_{n+1},
\end{gather*}
then (see~\cite{BM} or~\cite{BK})
\begin{gather*}
Y(v_j,z)=Y\big(\rho^{j-1}(v_1),z\big)=Y\big(v_1,e^{2(j-1)\pi i}z\big),
\\
Y\big(e^{v_j},z\big)=Y\big(e^{v_1},e^{2(j-1)\pi i}z\big),
\qquad
1<j\le n.
\end{gather*}
The factors $\frac{1}{\sqrt{2n}}$ and $\frac{1}{\sqrt{2}}$ in~\eqref{factors} follow from the fact that
$B_{v_1,-v_1}=4n$ and $B_{v_{n+1}, -v_{n+1}}=4$ (see~\cite[p.~853]{BM} for the def\/inition of these constants).
\end{Remark}
Let $\sigma$ be the isomorphism which sends $V$ to $F$, such that $\sigma(|0\rangle )=1$ and
$\sigma(|1\rangle )=\theta$,
\begin{gather}
\label{bfalpha}
\sigma\alpha^1_{-\frac{2j-1}{2n}-k}\sigma^{-1}=\frac{\hbar^{-\frac12}q^j_k}{\left({(2j-1)}/{(2n)}\right)_k},
\qquad
\sigma\alpha^1_{\frac{2j-1}{2n}+k}\sigma^{-1}=\left({(2j-1)}/{(2n)}\right)_{k+1}\hbar^{\frac12}
\frac{\partial}{\partial q^j_k},
\\[2mm]
\sigma\alpha^2_{-\frac{1}{2}-k}\sigma^{-1}=\frac{\hbar^{-\frac12}q^{n+1}_k}{\left({1}/{2}\right)_k},
\qquad
\sigma\alpha^2_{\frac{1}{2}+k}\sigma^{-1}=\left({1}/{2}\right)_{k+1}\hbar^{\frac12}\frac{\partial}
{\partial q^{n+1}_k},
\end{gather}
for $k=0,1,2,\ldots$ and $1\le j\le n$, where $(x)_k=x(x+1)\cdots (x+k-1)=\frac{\Gamma(x+k)}{\Gamma(x)}$ is
the (raising) Pochhammer symbol (N.B.\ $(x)_0=1$).
To describe $\sigma\phi^a(z)\sigma^-1$, we introduce two extra operators $\theta$ and
$\frac{\partial}{\partial \theta}$, then
\begin{gather*}
\sigma\phi^1(z)\sigma^{-1}=\frac{\left(\theta+\frac{\partial}{\partial\theta}\right)}{\sqrt2}z^{-\frac12}
\Gamma^1\big(q,z^{\frac1{2n}}\big),
\qquad
\sigma\phi^2(z)\sigma^{-1}=i\frac{\left(\theta-\frac{\partial}{\partial\theta}\right)}{\sqrt2}z^{-\frac12}
\Gamma^2\big(q,z^{\frac1{2}}\big),
\end{gather*}
where
\begin{gather}
\label{gggamma}
\Gamma^1\big(q,z^{\frac1{2n}}\big)=\Gamma^1_+\big(q,z^{\frac1{2n}}\big)\Gamma^1_-\big(q,z^{\frac1{2n}}\big),
\qquad
\Gamma^2\big(q,z^{\frac1{2}}\big)=\Gamma^2_+\big(q,z^{\frac1{2}}\big)\Gamma^2_-\big(q,z^{\frac1{2}}\big)
\end{gather}
with
\begin{gather}
\label{bf}
\Gamma^1_+\big(q,z^{\frac1{2n}}\big)=\exp\left(\frac{1}{\sqrt n}\sum_{j=1}^n\sum_{k=0}^\infty\frac{\hbar^{-\frac12}q^j_k}
{\left({(2j-1)}/{(2n)}\right)_{k+1}}z^{\frac{2j-1}{2n}+k}\right),
\\
\label{bf1}
\Gamma^1_-\big(q,z^{\frac1{2n}}\big)=\exp\left(\frac{1}{\sqrt n}\sum_{j=1}^n\sum_{k=0}^\infty-\left({(2j-1)}/{(2n)}
\right)_{k}\hbar^{\frac12}\frac{\partial}{\partial q^j_k}z^{-\frac{2j-1}{2n}-k}\right),
\\
\label{bf2}
\Gamma^2_+\big(q,z^{\frac12}\big)
=\exp\left(\sum_{k=0}^\infty\frac{\hbar^{-\frac12}q^{n+1}_k}{\left({1}/{2}\right)_{k+1}}z^{\frac{1}{2}+k}\right),
\\
\label{bf3}
\Gamma^2_-\big(q,z^{\frac12}\big)
=\exp\left(\sum_{k=0}^\infty-\left({1}/{2}\right)_{k}\hbar^{\frac12}\frac{\partial}{\partial q^{n+1}_k}z^{-\frac{1}{2}-k}\right).
\end{gather}
Now let $\sigma({\mathfrak T}_0)=\tau_0$ and $\sigma({\mathfrak T}_1)=\tau_1\theta$ Using~\eqref{bf}--\eqref{bf3} we can
rewrite the equation~\eqref{redDKP2} and thus obtain a~family of Hirota bilinear equations on $\tau_a$,
here $p\ge 0$:
\begin{gather}
\nonumber
\Res_\lambda\left(\lambda^{2np-1}\Gamma^1(q,\lambda)\tau_a\otimes\Gamma^1(q,-\lambda)\tau_b-(-)^{a+b}
\lambda^{2p-1}\Gamma^2(q,\lambda)\tau_a\otimes\Gamma^2(q,-\lambda)\tau_b\right)
\\
\qquad
=2\delta_{a+b,1}\delta_{p0}
\tau_b\otimes\tau_a.
\label{Hirotatau}
\end{gather}
{\it From now on we will often omit $\sigma$.}

Using Remark~\ref{r1}, we obtain that the total descendent potential of a~$D_{n+1}$ singularity
sa\-tisf\/ies~\eqref{Hirotatau}.

\section[Sato-Wilson and Lax equations]{Sato--Wilson and Lax equations}

\subsection{Pseudo-dif\/ferential operator approach}

We want to reformulate~\eqref{Hirotatau} in terms of pseudo-dif\/ferential operators.
For this we introduce an extra variable $x$ by replacing $q_0^1$ and $q_0^{n+1}$ by
$q_0^1+\frac{\hbar^{\frac12}}{2n}x$ and $q_0^{n+1}+\frac{\hbar^{\frac12}}{2}x$ and write $\partial$ for~$\partial_x$.
Then both $\tau_a$ and $\Gamma^b(q,\lambda)\tau_a$ for $b=1,2$, def\/ined in~\eqref{gggamma}, will depend
on $x$.
We keep the dependence in $\tau_a$ but remove it in the second term by writing
$\Gamma^b(x,q,\lambda)\tau_a=\Gamma^b(q,\lambda)\tau_a e^{x\lambda}$.
Next we rewrite~\eqref{redDKP2}:
\begin{gather*}
\Res_\lambda \Big(W(\lambda){\rm diag}\big(\lambda^{2np-1},\lambda^{2p-1}
\big)\otimes W(-\lambda)^T\Big)=\delta_{p0}V\otimes V^T,
\end{gather*}
where
\begin{gather}
W(\lambda)=
\begin{pmatrix}
\Gamma^1(q,\lambda)\tau_0&i\Gamma^2(q,\lambda)\tau_0
\\
i\Gamma^1(q,\lambda)\tau_1&\Gamma^2(q,\lambda)\tau_1
\end{pmatrix}
e^{x\lambda},
\qquad
V=
\begin{pmatrix}
\tau_1&i\tau_1
\\
i\tau_0&\tau_0
\end{pmatrix}
.
\end{gather}
Divide the f\/irst row of $W$ and $V$ by $\tau_1$ and the second by $\tau_0$, one thus obtains
\begin{gather}
\label{Hirotatau3}
\Res_\lambda \Big(P(\lambda){\rm diag}\big(\lambda^{2np-1},\lambda^{2p-1}\big)E(\lambda)e^{x\lambda}
\otimes e^{-x\lambda}E(-\lambda)^T P(-\lambda)^TJ\Big)=\delta_{p0}I,
\end{gather}
where
\begin{gather*}
P(\lambda)=\frac1{\sqrt2}
\begin{pmatrix}
\dfrac{\Gamma^1_-(q,\lambda)\tau_0}{\tau_1}&i\dfrac{\Gamma^2_-(q,\lambda)\tau_0}{\tau_1}
\vspace{1mm}\\
i\dfrac{\Gamma^1_-(q,\lambda)\tau_1}{\tau_0}&\dfrac{\Gamma^2_-(q,\lambda)\tau_1}{\tau_0}
\end{pmatrix}
\end{gather*}
and
\begin{gather*}
E(\lambda)=
\begin{pmatrix}
\Gamma^1_+(q,\lambda)&0
\\
0&\Gamma^2_+(q,\lambda)
\end{pmatrix},
\qquad
J=
\begin{pmatrix}
0&-i
\\
-i&0
\end{pmatrix}.
\end{gather*}
Then using the fundamental Lemma of~\cite{KL}, equation~\eqref{Hirotatau3} leads to:
\begin{gather*}
(P(\partial){\rm diag}\big(\partial^{2np-1},\partial^{2p-1}\big)P^*(\partial)J)_-=\delta_{p0}\partial^{-1}I.
\end{gather*}
Taking $p=0$ one deduces that
\begin{gather}
\label{Pinverse}
P^{-1}\partial^{-1}=\partial^{-1}P^*J
\end{gather}
and for $p>0$ that
\begin{gather*}
\big(P{\rm diag}\big(\partial^{2np},\partial^{2p}\big)P^{-1}\big)_{\le0}=0.
\end{gather*}
Now dif\/ferentiate~\eqref{Hirotatau3} for $p=0$ to some $q^j_k$ and apply the fundamental lemma then one
gets the following Sato--Wilson equations:
\begin{gather*}
\frac{\partial P}{\partial q^j_k}P^{-1}=-\big(B_k^j\big)_{\le 0},
\end{gather*}
where
\begin{gather*}
B_k^j=
\begin{cases}
\dfrac{1}{\sqrt n}\dfrac{\hbar^{-\frac12}}{{(2j-1)}/{(2n)})_{k+1}}PE_{11}\partial^{{2j-1}+{2kn}}
P^{-1}&\text{if}\quad j\le n,
\vspace{1mm}\\
\dfrac{\hbar^{-\frac12}}{({1}/{2})_{k+1}}PE_{22}\partial^{1+2k}P^{-1}&\text{if}\quad j=n+1.
\end{cases}
\end{gather*}
Now introduce the operators
\begin{gather*}
L=P\partial P^{-1},
\qquad
C_a=P E_{aa}P^{-1}.
\end{gather*}
Then clearly
\begin{gather}
[L,C_a]=0,
\qquad
C_aC_b=\delta_{ab}C_a,
\qquad
C_1+C_2=I,
\qquad
\big(L^{2np}C_1+L^{2p}C_2\big)_{\le0}=0\label{Ldiff}
\end{gather}
and one has the following Lax equations:
\begin{gather*}
\frac{\partial L}{\partial q^j_k}=\big[\big(B_k^j\big)_{>0},L\big],
\qquad
\frac{\partial C_a}{\partial q^j_k}=\big[\big(B_k^j\big)_{>0},C_a\big].
\end{gather*}
Note that in the important Drinfeld--Sokolov paper~\cite{DS}, in the case of the Coxeter element in the
Weyl group of type~D, also $2\times 2$ pseudo-dif\/ferential operators appear.
The principal realization of the basic representation is def\/initely related to this Drinfeld--Sokolov
hierarchy, see, e.g.,~\cite{FD}.
However, a~direct relation between our $2\times 2$ operators and the ones appearing in~\cite{DS} is unclear.

\subsection[The Orlov-Schulman and $S$ operator]{The Orlov--Schulman and $\boldsymbol{S}$ operator}

Introduce the Orlov--Schulman operator
\begin{gather*}
M=PE xE^{-1}P^{-1}=PRP^{-1},
\end{gather*}
where
\begin{gather*}
R=xI+2\hbar^{-\frac12}\sum_{k=0}^\infty\Bigg(\sqrt n E_{11}\sum_{j=1}^n\frac{q^j_k}{\left({(2j-1)}/{(2n)}
\right)_{k}}\partial^{2nk+{2j-2}}+E_{22}\frac{q^{n+1}_k}{\left(1/2\right)_{k}}\partial^{2k}\Bigg).
\end{gather*}
Then $[ L,M]=I$ and the wave function $W(\lambda)$ satisf\/ies
\begin{gather*}
LW(\lambda)=\lambda W(\lambda),
\qquad
C_iW(\lambda)=W(\lambda)E_{ii},
\qquad
MW(\lambda)=\frac{\partial W(\lambda)}{\partial\lambda}.
\end{gather*}
Moreover,
\begin{gather*}
M=\frac{\partial P}{\partial\partial}P^{-1}+2\hbar^{-\frac12}\sum_{k=0}^\infty\Bigg(\sqrt n\sum_{j=1}
^n\frac{q^j_k}{\left({(2j-1)}/{(2n)}\right)_{k}}L^{2nk+{2j-2}}C_1+\frac{q^{n+1}_k}{\left(1/2\right)_{k}}
L^{2k}C_2\Bigg).
\end{gather*}
We introduce the operator
\begin{gather*}
S=\left(\frac{1}{2n}ML^{1-2n}C_1+\frac12ML^{-1}C_2\right)_{\le0}P,
\end{gather*}
which will play a~crucial role in the deduction of the $W$ constraints.
$S$ is explicitly given by
\begin{gather}
S=\frac{1}{2n}\frac{\partial P}{\partial\partial}\partial^{1-2n}E_{11}+\frac{1}{2}\frac{\partial P}
{\partial\partial}\partial^{-1}E_{22}
\nonumber
\\
\phantom{S=}
{}+\sum_{k=0}^\infty\left(\frac{1}{\sqrt{n\hbar}}\sum_{j=1}^n\frac{q^{j}_k}{\left({(2j-1)}/{(2n)}\right)_{k}}
L^{2n(k-1)+{2j-1}}C_1+\frac1{\sqrt\hbar}\frac{q^{n+1}_k}{\left(1/2\right)_{k}}L^{2k-1}C_2\right)_{\le0}P
\nonumber
\\
\phantom{S}
=\frac{1}{2n}\frac{\partial P}{\partial\partial}\partial^{1-2n}E_{11}+\frac{1}{2}\frac{\partial P}
{\partial\partial}\partial^{-1}E_{22}+\frac{1}{\sqrt{n\hbar}}\sum_{j=1}^n{q^j_0}P\partial^{{2j-2n-1}}E_{11}
+\frac{1}{\sqrt\hbar}{q^{n+1}_0}P\partial^{-1}E_{22}
\nonumber
\\
\phantom{S=}
{}+\sum_{k=1}^\infty\left(\frac{1}{\sqrt{n\hbar}}\sum_{j=1}^n\frac{q^j_k}{\left({(2j-1)}/{(2n)}\right)_{k}}
L^{2n(k-1)+{2j-1}}C_1+\frac1{\sqrt\hbar}\frac{q^{n+1}_k}{\left(1/2\right)_{k}}L^{2k-1}C_2\right)_{\le0}P
\nonumber
\\
\phantom{S}
=\frac{1}{2n}\frac{\partial P}{\partial\partial}\partial^{1-2n}E_{11}+\frac{1}{2}\frac{\partial P}
{\partial\partial}\partial^{-1}E_{22}+\frac{1}{\sqrt{n\hbar}}\sum_{j=1}^n{q^j_0}P\partial^{{2j-2n-1}}E_{11}
+\frac1{\sqrt\hbar}{q^{n+1}_0}P\partial^{-1}E_{22}
\nonumber
\\
\phantom{S=}
-\sum_{j=1}^{n+1}\sum_{k=0}^\infty q^j_{k+1}\frac{\partial P}{\partial q^{j}_k}.
\label{S1}
\end{gather}

\section[The string equation and $W$ constraints]{The string equation and $\boldsymbol{W}$ constraints}

\subsection{The principal Virasoro algebra}

The principal realization of the basic representation of type $D_{n+1}^{(1)}$ has a~natural Virasoro
algebra with central charge $n+1$.
It is given by (see, e.g.,
\cite{tKvdL})
\begin{gather}
L_k=\sum_{j\in\mathbb{Z}}(-)^j\frac{j}{4n}:\phi_{\frac{-j}{2n}}^{1}\phi_{\frac{j}{2n}+k}^{1}:+(-)^j\frac{j}
{4}:\phi_{\frac{-j}{2}}^{2}\phi_{\frac{j}{2}+k}^{2}:+\delta_{k,0}\left(\frac{n+1}{16n}+\frac{n^2-1}{24n}
\right)
\nonumber
\\
\phantom{L_k}
=\sum_{j\in\mathbb{Z}}\frac{1}2:\alpha^1_{-\frac{1}{2n}-\frac{j}{n}}\alpha^1_{\frac{1}{2n}+\frac{j}{n}+k}
:+\frac{1}2:\alpha^2_{-\frac{1}{2}-j}\alpha^2_{\frac{1}{2}+j+k}:+\delta_{k,0}\left(\frac{n+1}{16n}
+\frac{n^2-1}{24n}\right),
\label{Vir}
\end{gather}
or in terms of the f\/ield
\begin{gather*}
L(z)=\sum\limits_{k\in\mathbb{Z}}L_k z^{-k-2}=\frac12w^{-\frac12}\frac{\partial}{\partial w}w^{\frac12}
\\
\phantom{L(z)=}
\times
\left.\left((-)^n:\phi^1(w)\phi^1\big(e^{2\pi i n}
z\big):-:\phi^2(w)\phi^2\big(e^{2\pi i}z\big):\right)\right|_{w=z}+\left(\frac{n+1}{16n}+\frac{n^2-1}{24n}\right)z^{-2}.
\end{gather*}
Using~\eqref{bfalpha} we can express $L_k$ in terms of the ``times" $q^j_k$, in particular $L_{-1}$ is
equal to
\begin{gather}
\label{Vir2}
\sigma L_{-1}\sigma^{-1}=\frac{1}{2\hbar}\left(q_0^{n+1}\right)^2+\frac{1}{2\hbar}\sum_{j=1}^n q_0^{j}
q_0^{n+1-j}+\sum_{\ell=1}^{n+1}\sum_{k=0}^\infty q_{k+1}^{\ell}\frac{\partial}{\partial q_{k}^{\ell}}.
\end{gather}
Let $\tau\in V_0$, the string equation is the following equation on $\tau$
\begin{gather}
\label{prestring}
L_{-1}\tau=\frac{\partial\tau}{\partial q_0^1}.
\end{gather}
However, following, e.g.,~\cite{FKN}, we remove the right-hand side of~\eqref{prestring} by introducing the shift $q_1^1\mapsto
q_1^1-1$.
This reduces the string equation to
\begin{gather}
\label{string}
L_{-1}\tau=0.
\end{gather}
However, this would introduce in the vertex operator $\Gamma^1_+(q,\lambda)$ of~\eqref{bf} some extra part
\begin{gather*}
e^{-\frac{(2n)^2\hbar^{-\frac12}}{\sqrt n}\frac{\lambda^{2n+1}}{2n+1}},
\end{gather*}
which fortunately cancels in~\eqref{Hirotatau}.
Therefor we will assume that the string equation is of the form~\eqref{string} and that the hierarchy is
given by~\eqref{Hirotatau}, where the operators~\eqref{bf} do not have this extra term.
We will show that if $\tau$ is in the $D_{n+1}^{(1)}$ group orbit of the vacuum vector, hence
satisf\/ies~\eqref{redDKP}, and $\tau$ satisf\/ies the string equation~\eqref{string}, i.e., that $\tau$ is
annihilated by $L_{-1}$, that this induces the annihilation of other elements in the $W_{D_{n+1}}$
$W$-algebra.
We will follow the approach of~\cite{vdL} (see also~\cite{vdL2}).
For this we use the following.
If $\tau=\tau_0=g| 0 \rangle$ satisf\/ies the string equation, then also its companion $\tau_1=g| 1
\rangle$, satisf\/ies the string equations.
This is because $\sigma L_{-1}\sigma^{-1}$ commutes with the operator $\theta+\frac{\partial}{\partial
\theta}$ which intertwines $F_0$ with $F_1$.

\subsection{A consequence of the string equation}

Assume that the string equation~\eqref{string} $L_{-1}\tau_a=0$ holds for both $a=0,1$.
Then clearly also
\begin{gather}
\label{string1}
\frac{\Gamma_-^c(\lambda)\left(L_{-1}\tau_a\right)}{\tau_b}-\frac{L_{-1}\tau_b}{(\tau_b)^2}
\Gamma_-^c(\lambda)(\tau_a)=0.
\end{gather}
Denote by $\tau^c_d=\Gamma_-^c(\lambda)(\tau_d)$, then $\eqref{string1}$ is equivalent to
\begin{gather}
\label{string2}
\frac{\tau_b\Gamma_-^c(\lambda)\left(L_{-1}\right)\tau_a^c-\tau_a^c L_{-1}\tau_b}{(\tau_b)^2}=0.
\end{gather}
Now,
\begin{gather*}
\Gamma_-^c(\lambda)\left(L_{-1}\right)=\frac{1}{2\hbar}\Gamma_-^c(\lambda)\left(\left(q_0^{n+1}
\right)^2+\sum_{j=1}^n q_0^{j}q_0^{n+1-j}\right)+\sum_{\ell=1}^{n+1}\sum_{k=0}
^\infty\Gamma_-^c(\lambda)\big(q_{k+1}^{\ell}\big)\frac{\partial}{\partial q_{k}^{\ell}},
\end{gather*}
hence~\eqref{string2} turns into
\begin{gather*}
\frac{1}{2\hbar}\frac{\tau_a^c}{\tau_b}\left(\Gamma_-^c(\lambda)-1\right)\left(\left(q_0^{n+1}
\right)^2+\sum_{j=1}^n q_0^{j}q_0^{n+1-j}\right)
\\
\qquad
+\sum_{\ell=1}^{n+1}\sum_{k=0}
^\infty\left(\frac{\Gamma_-^c(\lambda)\left(q_{k+1}^{\ell}\right)}{\tau_b}\frac{\partial\tau_a^c}
{\partial q_{k}^{\ell}}-\frac{\tau_a^c}{(\tau_b)^2}q_{k+1}^{\ell}\frac{\partial\tau_b}{\partial q_{k}^{\ell}
}\right)=0.
\end{gather*}
We rewrite this as
\begin{gather}
\label{string4}
\sum_{\ell=1}^{n+1}\sum_{k=0}^\infty q_{k+1}^{\ell}\frac{\partial\frac{\tau_a^c}{\tau_b}}{\partial q_{k}
^{\ell}}+R_{abc}=0,
\end{gather}
where
\begin{gather*}
R_{ab1}=\frac{\tau_a^1}{2\tau_b}\lambda^{-2n}-\frac{1}{\sqrt{n\hbar}}\frac{\tau_a^1}{\tau_b}\sum_{j=1}
^n\lambda^{1-2j}q_0^{n+1-j}
\\
\phantom{R_{ab1}=}
-\frac{\sqrt\hbar}{\sqrt n\tau_b}\sum_{\ell=1}^{n}\sum_{k=0}^\infty((2\ell-1)/2n)_{k+1}
\lambda^{1-2nk-2n-2\ell}\frac{\partial\tau_a^1}{\partial q_{k}^{\ell}}
\end{gather*}
and
\begin{gather*}
R_{ab2}=\frac{\tau_a^2}{2\tau_b}\lambda^{-2}-\frac{1}{\sqrt{\hbar}}\frac{\tau_a^2}{\tau_b}\lambda^{-1}
q_0^{n+1}-\frac{\sqrt\hbar}{\tau_b}\sum_{k=0}^\infty(1/2)_{k+1}\lambda^{-2k-3}\frac{\partial\tau_a^2}
{\partial q_{k}^{n+1}}.
\end{gather*}
We will now prove the following
\begin{Proposition}
The string equation~\eqref{string} induces
\begin{gather}
\label{string5}
\left(\left(\frac{1}{2n}ML^{1-2n}-\frac12L^{-2n}\right)C_1+\left(\frac12ML^{-1}-\frac12L^{-2}
\right)C_2\right)_{\le0}=0.
\end{gather}
\label{prop1}
\end{Proposition}
\begin{proof}
To prove this we f\/irst observe that~\eqref{string5} is equivalent to
\begin{gather}
\label{S2}
\left(\frac12P\partial^{-2n}E_{11}+\frac12P\partial^{-2}E_{22}\right)_{\le0}-S=0,
\end{gather}
where $S$ is given by~\eqref{S1}.
We calculate the various parts of this formula:
\begin{gather*}
\frac12P_{a1}\lambda^{-2n}-\frac{1}{\sqrt{n\hbar}}\sum_{j=1}^nq_0^jP_{a1}\lambda^{2j-2n-1}=\frac{i^{a-1}
\tau_{a-1}^1}{2\sqrt2\tau_{2-a}}\lambda^{-2n}-\frac{i^{a-1}\tau_{a-1}^1}{\sqrt{2n\hbar}\tau_{2-a}}\sum_{j=1}
^n\lambda^{2j-2n-1}q_0^j,
\\
\frac12P_{a2}\lambda^{-2}-\frac{1}{\sqrt{\hbar}}q_0^{n+1}P_{a2}\lambda^{-}=\frac{i(-i)^{a-1}\tau_{a-1}^1}
{2\sqrt2\tau_{2-a}}\lambda^{-2}-\frac{i(-i)^{a-1}\tau_{a-1}^1}{\sqrt{2\hbar}\tau_{2-a}}\lambda^{-1}.
\end{gather*}
Now
\begin{gather*}
\frac{1}{2n}\frac{\partial P_{a1}(\lambda)}{\partial\lambda}\lambda^{1-2n}=\frac{i^{a-1}\hbar^{\frac12}}
{\sqrt{2n}\tau_{2-a}}\frac{\partial\tau_{a-1}^1}{\partial\lambda}\lambda^{1-2n}
\\
\phantom{\frac{1}{2n}\frac{\partial P_{a1}(\lambda)}{\partial\lambda}\lambda^{1-2n}}
=\frac{i^{a-1}\hbar^{\frac12}}{\sqrt{2n}\tau_{2-a}}\sum_{\ell=1}^n\sum_{k=0}^\infty((2\ell-1)/2n)_{k+1}
\lambda^{1-2n(k+1)-2\ell}\frac{\partial\tau_{a-1}^1}{\partial q_k^\ell}
\end{gather*}
and
\begin{gather*}
\frac{1}{2}\frac{\partial P_{a2}(\lambda)}{\partial\lambda}\lambda^{-1}=\frac{i(-i)^{a-1}\hbar^{\frac12}}
{\sqrt2\tau_{2-a}}\frac{\partial\tau_{a-1}^1}{\partial\lambda}\lambda^{-1}
=\frac{i(-i)^{a-1}\hbar^{\frac12}}{\sqrt2\tau_{2-a}}\sum_{k=0}^\infty(1/2)_{k+1}\lambda^{-2k-3}
\frac{\partial\tau_{a-1}^1}{\partial q_k^{n+1}}.
\end{gather*}
Substituting these formulas into~\eqref{S2} one obtains up to a~multiplicative scalar
\begin{gather*}
\sum_{\ell=1}^{n+1}\sum_{k=0}^\infty
q^{\ell}_{k+1}\frac{\partial\tau_{a-1}^1/\tau_{2-a}}{\partial q_k^\ell}
+\frac{\tau_{a-1}^1}{2\tau_{2-a}}\lambda^{-2n}-\frac{\tau_{a-1}^1}{\sqrt{n\hbar}\tau_{2-a}}\sum_{j=1}
^n\lambda^{2j-2n-1}q_0^j
\\
\qquad
-\frac{\hbar^{\frac12}}{\sqrt n\tau_{2-a}}\sum_{\ell=1}^n\sum_{k=0}^\infty((2\ell-1)/2n)_{k+1}
\lambda^{1-2n(k+1)-2\ell}\frac{\partial\tau_{a-1}^1}{\partial q_k^\ell}=0
\end{gather*}
and
\begin{gather*}
\sum_{\ell=1}^{n+1}\sum_{k=0}^\infty q^{\ell}_{k+1}\frac{\partial\tau_{a-1}^1/\tau_{2-a}}{\partial q_k^\ell}
+\frac{\tau_{a-1}^1}{2\tau_{2-a}}\lambda^{-2}-\frac{\tau_{a-1}^1}{\sqrt{\hbar}\tau_{2-a}}\lambda^{-1}q_0^{n+1}
\\
\qquad
-\frac{\hbar^{\frac12}}{\tau_{2-a}}\sum_{k=0}^\infty(1/2)_{k+1}\lambda^{-2k+3}
\frac{\partial\tau_{a-1}^1}{\partial q_k^{n+1}}=0,
\end{gather*}
which is exactly equation~\eqref{string4}.
\end{proof}

A consequence of~\eqref{Ldiff} and Proposition~\ref{prop1}:
\begin{Proposition}
Let $\tau$ satisfy the string equation, then for all $p,q\ge 0$, except $p=q=0$, the following equation
holds:
\begin{gather}
\label{string6}
\left(\left(\frac{1}{2n}ML^{1-2n}-\frac12L^{-2n}\right)^q L^{2np}C_1+\left(\frac12ML^{-1}-\frac12L^{-2}
\right)^qL^{2p}C_2\right)_{\le0}=0.
\end{gather}
\end{Proposition}

We rewrite the formula~\eqref{string6}, using~\eqref{Pinverse}:
\begin{gather*}
\left(\left(\frac{1}{2n}PR\partial^{1-2n}-\frac12\partial^{-2n}\right)^q\partial^{2np-1}E_{11}
P^*J+\left(\frac12R\partial^{-1}-\frac12\partial^{-2}\right)^q\partial^{2p-1}E_{22}P^*J\right)_{-}=0.
\end{gather*}
Now using again the fundamental Lemma of~\cite{KL} this gives
\begin{gather}
\Res_\lambda\Bigg(\lambda^{2np-1}\left(\frac{1}{2n}\lambda^{1-2n}
\partial_\lambda-\frac12\lambda^{-2n}\right)^q W(\lambda)E_{11}
\\
\qquad
+\lambda^{2p-1}\left(\frac12\lambda^{-1}
\partial_\lambda-\frac12\lambda^{-2}\right)^qW(\lambda)E_{22}\Bigg)\otimes W(-\lambda)^T=0.\label{string8}
\end{gather}
Now let $\lambda^k=z$, then $\partial_z= \frac{1}{k}\lambda^{1-k}\partial_\lambda$ and
$\frac{1}{k}\lambda^{1-k}\partial_\lambda-\frac12\lambda^{-k}=z^{\frac12}\partial_z z^{-\frac12}$,
then~\eqref{string8} is equivalent~to
\begin{gather*}
\Res_z \Big(z^{p}\partial_z^q\left(z^{-\frac12}\Gamma^1\big(q,z^{\frac{1}{2n}}\big)\tau_a\right)
\otimes z^{-\frac12}\Gamma^1\big(q,-z^{\frac{1}{2n}}\big)\tau_b
\\
\qquad
{}-(-)^{a+b}z^{p}\partial_z^q\left(z^{-\frac12}\Gamma^2\big(q,z^{\frac{1}{2}}\big)\tau_a\right)\otimes z^{-\frac12}
\Gamma^2\big(q,-z^{\frac{1}{2}}\big)\tau_b\Big)=0.
\end{gather*}
And this formula induces
\begin{gather}
\Res_z\Big((-)^n z^{p}\partial_z^q\big(\phi^1(z)\big){\mathfrak T}_a\otimes\phi^1\big(e^{2\pi i n}
z\big){\mathfrak T}_b-z^{p}\partial_z^q\big(\phi^2(z)\big){\mathfrak T}_a\otimes\phi^2\big(e^{2\pi i}
z\big){\mathfrak T}_b\Big)=0,
\nonumber
\\
\Res_z\Big((-)^n z^{p}\phi^1\big(e^{2\pi i n}z\big){\mathfrak T}
_a\otimes\partial_z^q\big(\phi^1(z)\big){\mathfrak T}_b-z^{p}\phi^2\big(e^{2\pi i}z\big){\mathfrak T}
_a\otimes\partial_z^q\big(\phi^2(z)\big){\mathfrak T}_b\Big)=0.
\label{string10}
\end{gather}

\subsection{Some useful formulas}

We have
\begin{gather*}
\big[:\phi^1(y)\phi^1\big(e^{2\pi in}w\big):,\phi^1(z)\big]=\phi^1(y)\big\{\phi^1\big(e^{2\pi in}w\big),\phi^1(z)\big\}
-\big\{\phi^1(y),\phi^1(z)\big\}\phi^1\big(e^{2\pi in}w\big)
\\
\qquad
=(-)^n\sum_{j=0}^{2n-1}\delta_j(z-w)\phi^1(y)-\delta_j\big(z-e^{2\pi in}y\big)\phi^1\big(e^{2\pi in}w\big),
\end{gather*}
and similarly
\begin{gather*}
\big[:\phi^2(y)\phi^2\big(e^{2\pi i}w\big):,\phi^2(z)\big]
=-\sum_{j=0}^{1}\delta_{jn}(z-w)\phi^2(y)-\delta_{jn}\big(z-e^{2\pi i}y\big)\phi^2\big(e^{2\pi i}w\big).
\end{gather*}
We calculate the action of
\begin{gather*}
X(y,w)\otimes1=\left((-)^n:\phi^1(y)\phi^1\big(e^{2\pi in}w\big):-:\phi^2(y)\phi^2\big(e^{2\pi i}w\big):\right)\otimes1
\end{gather*}
on the bilinear identity~\eqref{redDKP2}, using the above formulas one obtains
\begin{gather*}
\delta_{a+b,1}\delta_{p0}X(y,w){\mathfrak T}_b\otimes{\mathfrak T}_a
\\
\quad
=\Res_zz^p\Bigg\{(-)^n\Bigg(\sum_{j=0}^{2n-1}\delta_j(z-w)\phi^1(y)-\delta_j\big(z-e^{2\pi in}y\big)
\phi^1\big(e^{2\pi in}w\big)\Bigg){\mathfrak T}_a\otimes\phi^{1}\big(e^{2\pi i n}z\big){\mathfrak T}_b
\\
\quad
\phantom{=}
-\Bigg(\sum_{j=0}^{1}\delta_{jn}(z-w)\phi^2(y)-\delta_{jn}\big(z-e^{2\pi i}y\big)\phi^2\big(e^{2\pi i}w\big)\Bigg)
{\mathfrak T}_a\otimes\phi^{1}\big(e^{2\pi i n}z\big){\mathfrak T}_b
\\
\quad
\phantom{=}
+(-)^n\phi^1(z)X(y,w){\mathfrak T}_a\otimes\phi^{1}(e^{2\pi i n}z){\mathfrak T}
_b-\phi^2(z)X(y,w){\mathfrak T}_a\otimes\phi^{2}(e^{2\pi i}z){\mathfrak T}_b\Bigg\}.
\end{gather*}
Thus
\begin{gather}
\delta_{a+b,1}\delta_{p0}X(y,w){\mathfrak T}_b\otimes{\mathfrak T}_a
\nonumber
\\
\quad\phantom{=}{}
-\Res_z \Big(z^p\left((-)^n\phi^1(z)X(y,w){\mathfrak T}_a\otimes\phi^{1}\big(e^{2\pi i n}z\big){\mathfrak T}
_b-\phi^2(z)X(y,w){\mathfrak T}_a\otimes\phi^{2}\big(e^{2\pi i}z\big){\mathfrak T}_b\right)\Big)
\nonumber
\\
\quad{}
=w^p\left((-)^n\phi^1(y){\mathfrak T}_a\otimes\phi^{1}\big(e^{2\pi i n}w\big){\mathfrak T}_b-\phi^2(y){\mathfrak T}
_a\otimes\phi^{2}\big(e^{2\pi i}w\big){\mathfrak T}_b\right)
\nonumber
\\
\quad\phantom{=}{}
-y^p\left((-)^n\phi^1\big(e^{2\pi i n}w\big){\mathfrak T}_a\otimes\phi^{1}(y){\mathfrak T}_b-\phi^2\big(e^{2\pi i}
w\big){\mathfrak T}_a\otimes\phi^{2}(y){\mathfrak T}_b\right).
\label{above}
\end{gather}

\subsection[$W$ constraints]{$\boldsymbol{W}$ constraints}

Now, let $X_{k\ell}=\Res_w w^k\partial_y^\ell X(y,w)|_{y=w}$, then putting $p=0$ in
formula~\eqref{above} and using~\eqref{string10}, one deduces
\begin{gather*}
\Res_z\Big((-)^n\phi^1(z)X_{pq}{\mathfrak T}_a\otimes\phi^{1}\big(e^{2\pi i n}z\big){\mathfrak T}
_b-\phi^2(z)X_{pq}{\mathfrak T}_a\otimes\phi^{2}\big(e^{2\pi i}z\big){\mathfrak T}_b\Big)\!=\delta_{a+b,1}X_{pq}
{\mathfrak T}_b\otimes{\mathfrak T}_a.
\end{gather*}
Thus
\begin{gather}
\Res_\lambda\lambda^{-1}\Big(\Gamma^1(q,\lambda)X_{pq}\tau_a\otimes\Gamma^1(q,-\lambda)
\tau_b-(-)^{a+b}\Gamma^2(q,\lambda)X_{pq}\tau_a\otimes\Gamma^2(q,-\lambda)\tau_b\Big)
\nonumber
\\
\qquad
=2\delta_{a+b,1}X_{pq}\tau_b\otimes\tau_a.
\label{stringtau}
\end{gather}
Note that here we abuse the notation, we write $X_{pq}$ for $\sigma X_{pq}\sigma^{-1}$.
Consider this as equation in two sets of variables~$x$,~$q$ and~$x'$,~$q'$.
Let $a\ne b$ and set $x=x'$ and $q=q'$, This gives
\begin{gather}
\label{taua=taub}
\frac{X_{pq}\tau_a}{\tau_a}=\frac{X_{pq}\tau_b}{\tau_b}.
\end{gather}
Now divide $\Gamma^c(q,\lambda)\left(X_{pq}\tau_a\right)$ by $\tau_b$, then
\begin{gather*}
\frac{\Gamma^c(q,\lambda)\left(X_{pq}\tau_a\right)}{\tau_b}
=\Gamma^c_+(q,\lambda)\frac{\Gamma^c_-(q,\lambda)\tau_a}{\tau_b}\Gamma^c_-(q,\lambda)\left(\frac{X_{pq}
\tau_a}{\tau_a}\right).
\end{gather*}
Now using~\eqref{taua=taub}, we rewrite~\eqref{stringtau} in the matrix version
\begin{gather*}
\Res_\lambda\left(\lambda^{-1}\sum_{c=1}^2\Gamma^c_-(q,\lambda)\left(\frac{X_{pq}\tau_a}{\tau_a}
\right)P(\lambda)E_{cc}E(\lambda)e^{x\lambda}\otimes e^{-x\lambda}E(-\lambda)^TP(-\lambda)^T J\right)=\frac{X_{pq}
\tau_a}{\tau_a}I.
\end{gather*}
Let
\begin{gather*}
\Gamma^c_-(q,\lambda)\left(\frac{X_{pq}\tau_a}{\tau_a}\right)=\sum_{k=0}^\infty S_k^c(x,q)\lambda^{-k},
\end{gather*}
then
\begin{gather*}
\Res_\lambda\left(\sum_{c=1}^2\sum_{k=0}^\infty S_k^c P(\lambda)E_{cc}\lambda^{-k-1}E(\lambda)e^{x\lambda}
\otimes e^{-x\lambda}E(-\lambda)^TP(-\lambda)^T J\right)=\frac{X_{pq}\tau_a}{\tau_a}I.
\end{gather*}
This gives
\begin{gather*}
\sum_{k=1}^\infty S_k^c P(\partial)E_{cc}\partial^{-k}P(\partial)^{-1}=0.
\end{gather*}
Now multiplying with $P(\partial)\partial^{\ell-1}$ from the right and taking the residue, one deduces that
\begin{gather*}
S_\ell^c(x,q)=0
\qquad
\text{for}
\quad
\ell=1,2,\dots,
\end{gather*}
hence,
\begin{gather*}
\left(\Gamma^c_-(q,\lambda)-1\right)\left(\frac{X_{pq}\tau_a}{\tau_a}\right)=0,
\end{gather*}
from which we conclude that
\begin{gather*}
\frac{X_{pq}\tau_a}{\tau_a}=\text{const}.
\end{gather*}
In order to calculate these constants, we determine $[X_{01}, X_{pq}]$ and $[X_{11}, X_{0q}]$.
The action of both operators on $\tau$ give zero.
Now write $X_{pq}=X_{pq}^1+X_{pq}^2$, then
\begin{gather*}
X^a_{pq}=\sum_{k>(q-p)n}(-)^k b_{pq}(k)\phi_{-\frac{k}{2n}}^a\phi_{\frac{k}{2n}+p-q}^a,
\end{gather*}
where
\begin{gather*}
b_{pq}(k)=\left(\frac{k}{2n}+\frac12-q\right)_q-\left(-\frac{k}{2n}+\frac12-p\right)_q.
\end{gather*}
From now on we assume $n=1$ if $a=2$,
in particular
\begin{gather*}
X_{01}^a=\sum_{k>n}(-)^k a(k)\phi_{-\frac{k}{2n}}^a\phi_{\frac{k}{2n}-1}^a,
\qquad
\text{where}
\qquad
a(k)=\frac{k}{n}-1.
\end{gather*}
Then
\begin{gather}
[X_{01}^a,X_{pq}^b]=\delta_{ab}\sum_{j>n,k>(q-p)n}(-)^k a(j)b_{pq}(k)\Big(\delta_{j-2n,k}\phi_{-\frac{j}
{2n}}^a\phi_{\frac{j}{2n}+p-q-1}^a
\nonumber
\\
\phantom{[X_{01}^a,X_{pq}^b]=}{}
+\delta_{j,-k+2(q-p+1)n}\phi_{-\frac{k}{2n}}^a\phi_{\frac{k}{2n}+p-q-1}^a-\delta_{j,k}\phi_{\frac{j}{2n}-1}
^a\phi_{-\frac{j}{2n}+p-q}^a
\nonumber
\\
\phantom{[X_{01}^a,X_{pq}^b]=}{}
-\delta_{j,k+2(p-q)n}\phi_{-\frac{k}{2n}}^a\phi_{\frac{k}{2n}+p-q-1}^a\Big).
\label{bereek}
\end{gather}
Now, if $p-q\ne 1$ the right hand side is normally ordered and we obtain
\begin{gather*}
[X_{01}^a,X_{pq}^b]=\delta_{ab}\sum_{j>(q-p+1)n}(-)^j\big(a(j)b_{pq}(j-2n)-a(j+2(p-q)n)b_{pq}
(j)\big)\phi_{-\frac{j}{2n}}^a\phi_{\frac{j}{2n}+p-q-1}^a.
\end{gather*}
It is straightforward to check that
\begin{gather*}
a(j)b_{pq}(j-2n)-a(j+2(p-q)n)b_{pq}(j)=-2pb_{p-1,q}(j),
\end{gather*}
thus
\begin{gather}
\label{bereek2}
\big[X_{01}^a,X_{pq}^b\big]=-2\delta_{ab}p X_{p-1,q}^b,
\qquad
\text{if}
\quad
p-q\ne1.
\end{gather}
If $p-q= 1$ we have to normal order the right hand side of~\eqref{bereek}.
Note that in that case, the second and third term of the right hand side of~\eqref{bereek} are equal to $0$
and the f\/irst term is normally ordered, the last one not.
This gives
\begin{gather*}
\big[X_{01}^a,X_{q+1,q}^b\big]=-2\delta_{ab}(q+1)X_{q,q}^b-2c_{q+1}^b,
\end{gather*}
where
\begin{gather*}
c_{q+1}^b=\left(\frac12a(2n)b_{q+1,q}(0)+\sum_{-n<k<0}a(k+2n)b_{q+1,q}(k)\right)
\\
\phantom{c_{q+1}^b}
=\sum_{j=1-n}^n\left(\frac{j}{2n}-q\right)_{q+1}+\left(-\frac{j}{2n}-q\right)_{q+1}.
\end{gather*}
Clearly, if $p=0$ the right hand side of~\eqref{bereek2} is equal to $0$.
For that case, one calculates
\begin{gather*}
[X_{11}^a,X_{0q}]=2q X_{0q},
\end{gather*}
so f\/inally we obtain the following result.
Note that $X_{p,0}=0$ and let $c_q=c_q^1+c_q^2$, then
\begin{Theorem}
\label{T1}
For all $p\ge 0$ and $q>0$, one has the following $W$ constraints:
\begin{gather*}
\left(X_{pq}+\frac{\delta_{p,q}}{2q+2}c_{q+1}\right)\tau_a=0,
\qquad
\text{for both}
\quad
a=0,1,
\end{gather*}
where
\begin{gather*}
c_q=\sum_{j=1-n}^n\left(\frac{j}{2n}-q\right)_{q}+\left(-\frac{j}{2n}-q\right)_{q}+\sum_{j=0}
^1\left(\frac{j}{2}-q\right)_{q}+\left(-\frac{j}{2}-q\right)_{q}.
\end{gather*}
\end{Theorem}
It is straightforward to check that for $|y|>|z|$
\begin{gather*}
X^a(y,z)=(-)^n:\phi^a(y)\phi^a(e^{2\pi i n}z):
\\
\phantom{X^a(y,z)}
=\frac12(yz)^{-\frac12}\frac{y^{\frac{1}{2n}}+z^{\frac{1}{2n}}}{y^{\frac{1}{2n}}-z^{\frac{1}{2n}}}
\left(\Gamma_+^a\big(q,y^{\frac{1}{2n}}\big)\Gamma_+^a\big(q,-z^{\frac{1}{2n}}\big)
\Gamma_-^a\big(q,y^{\frac{1}{2n}}\big)\Gamma_-^a\big(q,-z^{\frac{1}{2n}}\big)-1\right)
\end{gather*}
and
\begin{gather}
\label{XXX}
X_{pq}^a=\frac{(-1)^n}{q+1}\Res_zz^p\partial_y^{q+1}(y-z):\phi^a(y)\phi^a\big(e^{2\pi i n}z\big):\bigg|_{y=z}.
\end{gather}

Now
\begin{gather*}
\partial_y^k\left((y-z)(yz)^{-\frac12}\frac{y^{\frac{1}{2n}}+z^{\frac{1}{2n}}}{y^{\frac{1}{2n}}-z^{\frac{1}
{2n}}}\right)\Bigg|_{y=z}=c^a_kz^{-k}.
\end{gather*}
Let
\begin{gather*}
\Gamma^a(y,z)=\Gamma_+^a\big(q,y^{\frac{1}{2n}}\big)\Gamma_+^a\big(q,-z^{\frac{1}{2n}}\big)
\Gamma_-^a\big(q,y^{\frac{1}{2n}}\big)\Gamma_-^a\big(q,-z^{\frac{1}{2n}}\big),
\end{gather*}
then
\begin{gather*}
X_{pq}^a=\Res_z\frac{1}{2q+2}\sum_{k=0}^{q+1}
\begin{pmatrix}
q+1
\\
k
\end{pmatrix}
c_k^a z^{p-k}\partial_y^{q-k+1}\left(\Gamma^a(y,z)-1\right)\Bigg|_{y=z}
\end{gather*}
and
\begin{gather*}
X_{pq}^a+\frac{\delta_{p,q}}{2q+2}c_{q+1}^a=\Res_z\frac{1}{2q+2}\sum_{k=0}^{q+1}
\begin{pmatrix}
q+1
\\
k
\end{pmatrix}
c_k^a z^{p-k}\partial_y^{q-k+1}\left(\Gamma^a(y,z)\right)\Bigg|_{y=z}.
\end{gather*}
We now want to obtain one formula in which we combine all our $W$ constraints.
For this, we f\/irst write the generating series of the $c_k^a$:
\begin{gather*}
\sum_{k=0}^\infty\frac{c_k^a}{k!}z^k=\sum_{k=0}^\infty\left(\sum_{j=1-n}^n
\begin{pmatrix}
\frac{j}{2n}
\\
k
\end{pmatrix}
+
\begin{pmatrix}
-\frac{j}{2n}
\\
k
\end{pmatrix}
\right)z^k
=\sum_{j=1-n}^n\left((1+z)^{\frac{j}{2n}}+(1+z)^{-\frac{j}{2n}}\right).
\end{gather*}
Next we calculate for $|u|>|z|>|w|$,
\begin{gather*}
\sum_{p,q=0}^\infty\frac{X_{pq}^a}{q!}u^{-p-1}w^{q+1}
\\
\qquad
=\frac12\sum_{p,q=0}^\infty\Res_z\frac{u^{-p-1}w^{q+1}}{(q+1)!}\sum_{k=0}^{q+1}
\begin{pmatrix}
q+1
\\
k
\end{pmatrix}
c_k^a z^{p-k}\left(\partial_y\right)^{q-k+1}\left(\Gamma^a(y,z)-1\right)\Bigg|_{y=z}
\\
\qquad
=\frac12\Res_z\frac{1}{u-z}\sum_{q=0}^\infty\sum_{k=0}^{q+1}\frac{c_k^a}{k!}\left(\frac{w}{z}
\right)^k\frac{\left(w\partial_y\right)^{q-k+1}}{(q-k+1)!}\left(\Gamma^a(y,z)-1\right)\Bigg|_{y=z}
\\
\qquad
=\frac12\Res_z\frac{1}{u-z}\sum_{k=0}^\infty\sum_{\ell=0}^{\infty}\frac{c_k^a}{k!}\left(\frac{w}{z}
\right)^k\frac{\left(w\partial_y\right)^{\ell}}{\ell!}\left(\Gamma^a(y,z)-1\right)\Bigg|_{y=z}
\\
\qquad
=\frac12\Res_z\frac{1}{u-z}\sum_{j=1-n}^n\left(\left(1+\frac{w}{z}\right)^{\frac{j}{2n}}
+\left(1+\frac{w}{z}\right)^{-\frac{j}{2n}}\right)\left(\Gamma^a(z+w,z)-1\right).
\end{gather*}
Note that
\begin{gather*}
\Res_z\frac{1}{u-z}\sum_{j=1-n}^n\left(\left(1+\frac{w}{z}\right)^{\frac{j}{2n}}
+\left(1+\frac{w}{z}\right)^{-\frac{j}{2n}}\right)
\\
\qquad{}
=\sum_{j=1-n}^n\left(\left(1+\frac{w}{u}\right)^{\frac{j}{2n}}+\left(1+\frac{w}{u}\right)^{-\frac{j}{2n}}\right)-c_0^a.
\end{gather*}
Thus we have
\begin{Theorem}
\label{T2}
For $|u|>|z|>|w|$, one has the following $W$ constraints:
\begin{gather*}
\Res_z\frac{1}{u-z}
\left(\sum_{j=1-n}^n\left(\left(1+\frac{w}{z}\right)^{\frac{j}{2n}}+\left(1+\frac{w}{z}\right)^{-\frac{j}{2n}}\right)
\Gamma^1(z+w,z) \right.
\\
\left.\qquad{}
+
\sum_{j=0}^1\left(\left(1+\frac{w}{z}\right)^{\frac{j}{2}}+\left(1+\frac{w}{z}\right)^{-\frac{j}{2}}
\right)\Gamma^2(z+w,z)\right)\tau_a=0.
\end{gather*}
\end{Theorem}

We can express this in a~dif\/ferent manner.
Def\/ine
\begin{gather*}
q^a_1[z]=\partial_y\Gamma^a(y,z)\big|_{y=z}
\qquad
\text{and}
\qquad
q^a_r[z]=\partial_z^{r-1}q^a_1[z],
\end{gather*}
then
\begin{gather*}
q^1_r[z]=\frac{1}{\sqrt n}\sum_{j=1}^n\sum_{k=0}^\infty\Bigg(\frac{\hbar^{-\frac12}q^j_k}{\left({(2j-1)}
/{(2n)}\right)_{k+1-r}}z^{\frac{2j-1}{2n}+k-r}
\\
\phantom{q^1_r[z]=}
-\left({(2j-1)}/{(2n)}\right)_{k+r}\hbar^{\frac12}
\frac{\partial}{\partial q^j_k}z^{-\frac{2j-1}{2n}-k-r}\Bigg),
\\
q^2_r[z]=\sum_{k=0}^\infty\left(\frac{\hbar^{-\frac12}q^{n+1}_k}{\left({1}/{2}\right)_{k+1-r}}z^{\frac{1}
{2}+k-r}-\left({1}/{2}\right)_{k+r}\hbar^{\frac12}\frac{\partial}{\partial q^{n+1}_{k}}z^{-\frac{1}{2}-k-r}
\right),
\end{gather*}
here we use the convention that for $m>0$
\begin{gather*}
\frac{1}{(a)_{-m}}=\frac{\Gamma(a)}{\Gamma(a-m)}=(a-m)_m.
\end{gather*}
Thus
\begin{gather*}
\frac{X_{pq}^a}{q!}+\frac{\delta_{pq}}2\frac{c_{q+1}}{(q+1)!}=\frac12\Res_z\sum_{\ell=0}^{q+1}
z^{p-\ell}\frac{c_{\ell}}{\ell!}:S_{q-\ell+1}\left(\frac{q^a_r[z]}{r!}\right):,
\end{gather*}
where the $S_\ell(x)$ are the elementary Schur functions def\/ined by
\begin{gather*}
\sum_{\ell=0}^\infty S_\ell(x)=\exp\left(\sum_{k=1}^\infty x_kz^k\right).
\end{gather*}
Thus we have the following consequence of Theorem~\ref{T2}:
\begin{Corollary}
For $|u|>|z|>|w|$,
\begin{gather*}
\Res_z\frac{1}{u-z}
\Bigg(\sum_{j=1-n}^n\left(\left(1+\frac{w}{z}\right)^{\frac{j}{2n}}+\left(1+\frac{w}{z}
\right)^{-\frac{j}{2n}}\right):e^{\sum\limits_{r=1}^\infty\frac{q^1_r[z]w^r}{r!}}:
\\
\qquad
{}+\sum_{j=0}^1\left(\left(1+\frac{w}{z}\right)^{\frac{j}{2}}
+\left(1+\frac{w}{z}\right)^{-\frac{j}{2}}\right):e^{\sum\limits_{r=1}^\infty\frac{q^2_r[z]w^r}{r!}}:\Bigg)\tau_a=0.
\end{gather*}
\end{Corollary}
A similar result is described in~\cite[Section~3.5]{BM}.

\section{A comparison with the results of Bakalov and Milanov~\cite{BM}}

Unfortunately we do not obtain all the $W$ constraints of Bakalov and Milanov~\cite{BM} from the string
equation.
Kac, Wang and Yan gave a~description in~\cite{KWY} of the corresponding $W$ algebra.
As is mentioned in~\cite[Example 2.5]{BM}, this $W$ algebra is generated by the elements (cf.\
Remark~\ref{relBM})
\begin{gather*}
\nu^d:=\sum_{i=1}^{n+1}e^{v_i}\,_{(-d)}e^{-v_i}+e^{-v_i}\,_{(-d)}e^{v_i}
\\
\phantom{\nu^d\,}
=\sum_{i=1}^{2n}e^{\rho^i(v_1)}\,_{(-d)}e^{\rho^{n+i}(v_1)}+\sum_{i=1}^{2}e^{\rho^i(v_{n+1})}\,_{(-d)}
e^{\rho^{i+1}(v_{n+1})},
\qquad
d>0,
\end{gather*}
and the element
\begin{gather*}
\pi^{n+1}:={v_{1}}_{(-1)}{v_{2}}_{(-1)}\cdots_{(-1)}{v_{n}}_{(-1)}v_{n+1}.
\end{gather*}
Our constraints come from the elements $\nu^d$, the constraints related to the element $\pi^{n+1}$ cannot
be obtained from the string equation.

Since the total descendent potential is a~highest weight vector of this W algebra, this means (Theorem 1.1
of~\cite{BM}) that it is annihilated by all coef\/f\/icients of the fractional powers of $z$, where the
power is $\le -1$, of all $Y(\nu^d,z)$ and $Y(\pi^{n+1},z)$.
Now
\begin{gather*}
Y\big(\nu^d,z\big)=\frac{1}{(d+1)!}\partial_{y}^{d+1}(y-z)
\\
\qquad
\times
\Bigg(\sum_{j=1}^{2n}Y\big(e^{\rho^j(v_1)},y\big) Y\big(e^{\rho^{j+n}(v_1)},z\big)
+\sum_{j=1}^{2}Y\big(e^{\rho^j(v_{n+1})},y\big) Y\big(e^{\rho^{j+1}(v_{n+1})},z\big)\Bigg)\Bigg|_{y=z}.
\end{gather*}
Using Remark~\ref{relBM} we obtain that
\begin{gather}
Y\big(\nu^d,z\big)=\frac{1}{(d+1)!}\partial_{y}^{d+1}(y-z)
\nonumber
\\
\qquad
\times
\Bigg(\frac{(-)^n}{{2n}}\sum_{j=1}^{2n}\phi^1\big(e^{2j\pi i}y\big)\phi^1\big(e^{2(j+n)\pi i}z\big)
-\frac{1}{2}\sum_{j=1}^{2}\phi^2\big(e^{2j\pi i}y\big)\phi^2\big(e^{2(j+1)\pi i}z\big)\Bigg)\Bigg|_{y=z}.
\label{YYY}
\end{gather}
Using the fact that $1+\omega +\omega^2+\dots +\omega^{k-1}=0$ for $\omega\ne 1$ a~$k$-th root of~1, one
obtains that all non-integer powers of $z$ do not appear in~\eqref{YYY}.
Hence,
\begin{gather*}
Y\big(\nu^d,z\big)\!=\Res_w\delta(z\!-\!w)\frac{1}{(d\!+\!1)!}\partial_{y}^{d+1}
(y\!-\!w)\left((-)^n\phi^1(y)\phi^1\big(e^{2n\pi i}w\big)-\phi^2(y)\phi^2\big(e^{2\pi i}w\big)\right)\!\bigg|_{y=w}\!.
\end{gather*}
Using~\eqref{XXX}, we see that the total descendent potential gets annihilated by
\begin{gather*}
X_{pq}+\frac{\delta_{p,d}}{2q+2}c_{d+1},
\qquad
p\ge0, \qquad  d=1,2,\ldots,
\end{gather*}
which are exactly the constraints appearing in Theorem~\ref{T1}.

\section{The string equation on the Grassmannian}

Using the~\eqref{Vir}-formulation of $L_{-1}$ in terms of the elements $\phi_i^a$, one can show that
\begin{gather*}
\Big[L_{-1},\phi_{\frac{k}{2n}}^{1}\Big]=\left(\frac12-\frac{k}{2n}\right)\phi_{\frac{k}{2n}-1}^{1},
\qquad
\Big[L_{-1},\phi_{\frac{k}{2}}^{2}\Big]=\left(\frac12-\frac{k}{2}\right)\phi_{\frac{k}{2}-1}^{1}.
\end{gather*}
Then using the identif\/ication~\eqref{phit} we obtain
\begin{gather*}
\Big[L_{-1},t^{\frac{k}{2n}}e_1\Big]=-\bigg(t^{\frac12}\frac{d}{dt}t^{-\frac12}\bigg)\big(t^{\frac{k}{2n}}\big)e_1,
\qquad
\Big[L_{-1},t^{\frac{k}{2}}e_2\Big]=-\bigg(t^{\frac12}\frac{d}{dt}t^{-\frac12}\bigg)\big(t^{\frac{k}{2}}\big)e_2.
\end{gather*}
Now applying the dilaton shift $q_1^1\mapsto q_1^1+1$, then $\sigma L_{1}\sigma^{-1}$
changes according to the description~\eqref{Vir2} to
\begin{gather*}
\sigma L_{1}\sigma^{-1}+\frac{\partial}{\partial q_0^1},
\end{gather*}
and by~\eqref{bfalpha} one f\/inds that $L_{-1}$ changes into
$L_{-1}+2n\hbar^{-\frac12}\alpha_{\frac{1}{2n}}^1$.
Since
\begin{gather*}
\Big[\alpha_{\frac{1}{2n}}^1,\phi^a(z)\Big]=\frac{\delta_{a1}}{\sqrt n}z^{\frac{1}{2n}}\phi^1(z),
\end{gather*}
we obtain
\begin{Proposition}
Let $W$ be the point of the Grassmannian which corresponds to the $\tau$-function
that satisfies the string equation, then $W$ satisfies
\begin{gather*}
tW\subset W
\qquad
\text{and}
\qquad
\left(-\frac{d}{dt}+\frac12t^{-1}+2\sqrt n\hbar^{-\frac12}t^{\frac1{2n}}E_{11}\right)W\subset W.
\end{gather*}
\end{Proposition}
Note that the total descendent potential of a~$D_{n+1}$ type singularity is tau function, that satisf\/ies
this condition.
Vakulenko, used a~similar approach in~\cite{V}.
He showed that the tau function is unique.
However, his action on the Grassmannian seems somewhat strange.

\subsubsection*{Acknowledgements}

I would like to thank Bojko Bakalov for useful discussions and the three referees for valuable suggestions,
which improved the paper.

\pdfbookmark[1]{References}{ref}
\LastPageEnding

\end{document}